\documentclass[12pt]{article} 

\usepackage[utf8]{inputenc}
\usepackage{naist-mthesis} 

\usepackage[usenames,dvipsnames,svgnames,table]{xcolor}
\usepackage{comment}

\usepackage{cite}
\usepackage{graphicx}
\usepackage{textcomp}
\usepackage{url}
\usepackage{CJKutf8}
\usepackage{multirow}

\newcommand{\bm}[1]{\text{\boldmath $#1$}}

\newcommand{\X}{\mathcal{X}}

\usepackage{caption}
\usepackage{subcaption}
\usepackage[fleqn]{amsmath}
\usepackage{nccmath}
\usepackage{latexsym}
\usepackage[psamsfonts]{amssymb}
\usepackage{amsfonts}
\usepackage{amsthm}
\usepackage{algorithm}
\usepackage[noend]{algpseudocode}
\usepackage{filecontents}
\usepackage{csvsimple}
\usepackage{autobreak}
\usepackage{longtable}

\studentnumber{2011301}
\elabname{Large-Scale Systems Management Lab.}

\doctitle{\mastersthesis}      
\major{\engineering}    
\program{\ise}         

\title{Dealing with Imbalanced Classes in Bot-IoT Dataset}
\ptitle{}
\etitle{Dealing with Imbalanced Classes in Bot-IoT Dataset}
\eptitle{}

\author{}
\pauthor{}
\eauthor{Jesse Atuhurra}
\epauthor{}

\esyear{2022}
\smonth{3}
\sday{17}

\ecmembers{Professor Shoji Kasahara}{(Supervisor, Division of Information Science)}
                        {Professor Youki Kadobayashi}{(Co-Supervisor, Division of Information Science)}
                        {Associate Professor Masahiro Sasabe}{(Co-Supervisor, Division of Information Science)}
                        {Associate Professor YuanYu Zhang}{(Co-Supervisor, Xidian University)}

\eaddcmembers{Assistant Professor Takanori Hara}{(Co-Supervisor, Division of Information Science)}
           {}{}
           {}{}
           {}{}

\pkeywords{}
\ekeywords{Network intrusion detection system (NIDS), IoT network, SMOTE algorithm, class imbalance problem, machine learning}
\epkeywords{}

\abstract{
With the rapidly spreading usage of internet of things (IoT) devices, a network intrusion detection system (NIDS) has an important role to detect and protect various types of attacks in the IoT network.
To evaluate the robustness of the NIDS in the IoT network, the existing work proposed a realistic botnet dataset in the IoT network (Bot-IoT dataset) and applied it to machine learning based anomaly detection.
This dataset contains imbalanced normal and attack packets because the number of normal packets is much smaller than that of attack ones.
The nature of imbalanced data may make it difficult to identify the
minority class correctly.
In this thesis, to address the class imbalance problem in the Bot-IoT dataset, we propose a binary classification method with synthetic minority over-sampling techniques (SMOTE).
The proposed classifier aims at detecting the attack packets and overcoming the class imbalance problem with the help of SMOTE algorithm.
Through numerical results, we demonstrate the fundamental characteristics of the proposed classifier and the impact of imbalanced data on the classifiers' performance.
}
\pabstract{ }
\eabstract{
With the rapidly spreading usage of internet of things (IoT) devices, a network intrusion detection system (NIDS) has an important role to detect and protect various types of attacks in the IoT network.
To evaluate the robustness of the NIDS in the IoT network, the existing work proposed a realistic botnet dataset in the IoT network (Bot-IoT dataset) and applied it to machine learning based anomaly detection.
This dataset contains imbalanced normal and attack packets because the number of normal packets is much smaller than that of attack ones.
The nature of imbalanced data may make it difficult to identify the
minority class correctly.
In this thesis, to address the class imbalance problem in the Bot-IoT dataset, we propose a binary classification method with synthetic minority over-sampling techniques (SMOTE).
The proposed classifier aims at detecting the attack packets and overcoming the class imbalance problem with the help of SMOTE algorithm.
Through numerical results, we demonstrate the fundamental characteristics of the proposed classifier and the impact of imbalanced data on the classifiers' performance.
}

\epabstract{ }

\begin{document}
\titlepage
\cmemberspage
\firstabstract

\toc
\newpage
\listoffigures
\newpage
\listoftables
\pagenumbering{arabic}

\newpage
\section{Introduction}
\label{sec:Introduction}Nowadays, the internet of things (IoT) plays a key role to facilitate and advance services and encompasses various application domains~\cite{khurpadeSurveyIOT5G2018}.
With rapidly increasing heterogeneous connected devices, that is, IoT devices, various services are constantly being created, and thus the network traffic in IoT networks has exponentially been increasing.
To protect these devices against cyber-attacks included in the huge amount of network traffic, the need for IoT security arises~\cite{hassijaSurveyIoTSecurity2019}.
The existing IoT networks are vulnerable to various types of attacks due to the accessibility to devices from anywhere via the internet and the proliferation of low-level security protection.
These attacks lead to not only critical damage of infrastructures but also privacy violation.
Since the distributed nature of IoT networks makes it difficult to monitor the network and to collect audit data, it is hard to take an effective security strategy against various attacks included in the huge amount of network traffic due to the heterogeneity of IoT devices and the limited resources available in IoT devices.

Under such background, IoT networks have increasingly become susceptible to intruder attacks, e.g., denial of service (DoS), distributed denial of service (DDoS), probe attacks~, and IP spoofing.
To protect these attacks by intruders, an intrusion detection system (IDS)~\cite{liaoIntrusionDetectionSystem2013, khraisatSurveyIntrusionDetection2019} was proposed, which is a monitoring system to detect abnormal and/or malicious activities by analyzing audit data.
The existing studies have addressed the anomaly detection using the IDS, which monitors the normal activities of network traffic and alerts the administrator to the activity if any activities which deviate from the normal activities are found.
In addition, several studies have proposed a network-based IDS (NIDS)~\cite{ahmedSurveyNetworkAnomaly2016} placed at the fog node in the IoT network to address the issue of huge latency~\cite{aliyuDetectionPreventionTechnique2018, saharDeepLearningApproachBased2021, reddyExactGreedyAlgorithm2021, chaabouni2019network}.

The NIDS's built on machine learning (ML) have attracted many researchers~\cite{chaabouni2019network, saharDeepLearningApproachBased2021, gulghaneSurveyIntrusionDetection2020, rahmanScalableMachineLearningbased2020, deorankarSurveyAnomalyDetection2020, thakkarReviewMachineLearning2021}, since the existing NIDS has the limitation of scalability and autonomic self-adoption due to constantly evolving attacks and threats in the IoT network.
The combination of an NIDS and ML is expected to overcome these limitations.
However, the existing ML methods, e.g., support vector machine (SVM), $k$-nearest neighbor (KNN), and decision tree (DT), may demonstrate low performance on anomaly detection because the NIDS requires high-dimensional representation.
Deep neural network based NIDS's were proposed~\cite{kwonSurveyDeepLearningbased2019, ahmadNetworkIntrusionDetection2021}, which can exhibit high performance by exploring the high-dimensional representation from the audit data.
In general, the audit data generated from network traffic tends to have the class imbalance problem because the activity which deviates from the normal activities, i.e., the attack, is a rare case.
More precisely, such imbalanced traffic data has majority of normal traffic and minority of abnormal traffic.
With imbalanced data, it becomes difficult for the classifier to correctly identify traffic which belongs to the minority class.

To apply ML to the NIDS, there are various datasets related to the cyber-attacks~\cite{koroniotis2019towards, sharafaldinTowardGenerating2018, tavallaeeDetailedAnalysisKDD2009, moustafaUNSWNB15ComprehensiveData2015, shiraviDevelopingSystematicApproach2012, KDDCup1999Data}.
In particular, Koroniotis et al.\ published the realistic botnet dataset, for network intrusion detection and network forensic analytics, which contains IoT network traffic including various types of attacks~\cite{koroniotis2019towards}.
In \cite{koroniotis2019towards}, they evaluated the reliability of this dataset by using several ML methods and showed the baseline of binary classification.
This dataset also contains imbalanced normal and attack packets because the number of normal packets is much smaller than that of attack ones.

We will describe the detail of the Bot-IoT dataset in Section~\ref{sec:Bot-IoT Dataset}.

In this thesis, to address the class imbalance problem in the Bot-IoT dataset, we propose a binary classification method at the network packet level considering the nature of imbalanced data.
The proposed method aims at identifying normal or attack packets and overcoming the class imbalance problem with the help of the existing sampling method, i.e., synthetic minority over-sampling technique (SMOTE)~\cite{Chawla_2002}. Before selecting SMOTE technique for sampling, we report that we tried random over sampling and random under sampling too but SMOTE sampling showed more consistent results. Therefore we used SMOTE sampling for further analysis.
Through numerical results, we demonstrate the fundamental characteristics of the proposed method from the viewpoint of the impact of SMOTE sampling on the performance of classifiers, during intrusion detection. Moreover, though binary classification is reported in this work, it is possible to extend the binary classification to multi-class classification especially after an attack has been detected, to identify the actual category of the attack.

The main contributions of this thesis are as follows: 
\begin{enumerate}
  \item Use the Bot-IoT dataset to develop a NIDS for IoT networks.
  \item Investigate the class imbalance problem in Bot-IoT dataset.
\end{enumerate}
The rest of this thesis is organized as follows.
Section~\ref{sec:Related Work} gives the related work.
In Section~\ref{sec:Background}, we introduce the Bot-IoT dataset and ML techniques used in this thesis.
In Section~\ref{sec:Proposed Method}, we propose a binary classification method to deal with the class imbalance.
Section~\ref{sec:Numerical Results} shows the fundamental characteristics of the proposed method.
Finally, we give the conclusion and future work in Section~\ref{sec:Conclusion}.
\newpage

\section{Related Work}
\label{sec:Related Work}

Many researchers have studied an NIDS to protect various types of attacks by intruders~\cite{ahmedSurveyNetworkAnomaly2016}.
An NIDS runs on a strategic point and inspects the traffic among all the devices on the network.
In the domain of IoT networks, several studies have proposed NIDS's placed at the fog node in the IoT network~\cite{aliyuDetectionPreventionTechnique2018, reddyExactGreedyAlgorithm2021, saharDeepLearningApproachBased2021}.
Aliyu et al.\ proposed a resource efficient IDS for man in the middle attacks at the fog layer~\cite{aliyuDetectionPreventionTechnique2018}.
There are studies to apply ML to the NIDS~\cite{reddyExactGreedyAlgorithm2021, saharDeepLearningApproachBased2021}.
Reddy et al.\ proposed an extreme greedy boosting ensemble method~\cite{chenXGBoostScalableTree2016} based NIDS running at the fog node to monitor network traffic in IoT network by identifying and classifying the attacks based on abnormal activities~\cite{reddyExactGreedyAlgorithm2021}.
Sahar et al.\ developed a deep learning based NIDS implemented on the fog node~\cite{saharDeepLearningApproachBased2021}.
Recent surveys can be found in \cite{chaabouni2019network, gulghaneSurveyIntrusionDetection2020, rahmanScalableMachineLearningbased2020, deorankarSurveyAnomalyDetection2020, thakkarReviewMachineLearning2021}.
In this thesis, we propose the ML-based binary classification for an NIDS in IoT network.

\begin{figure}[t]
\includegraphics[width=\columnwidth]{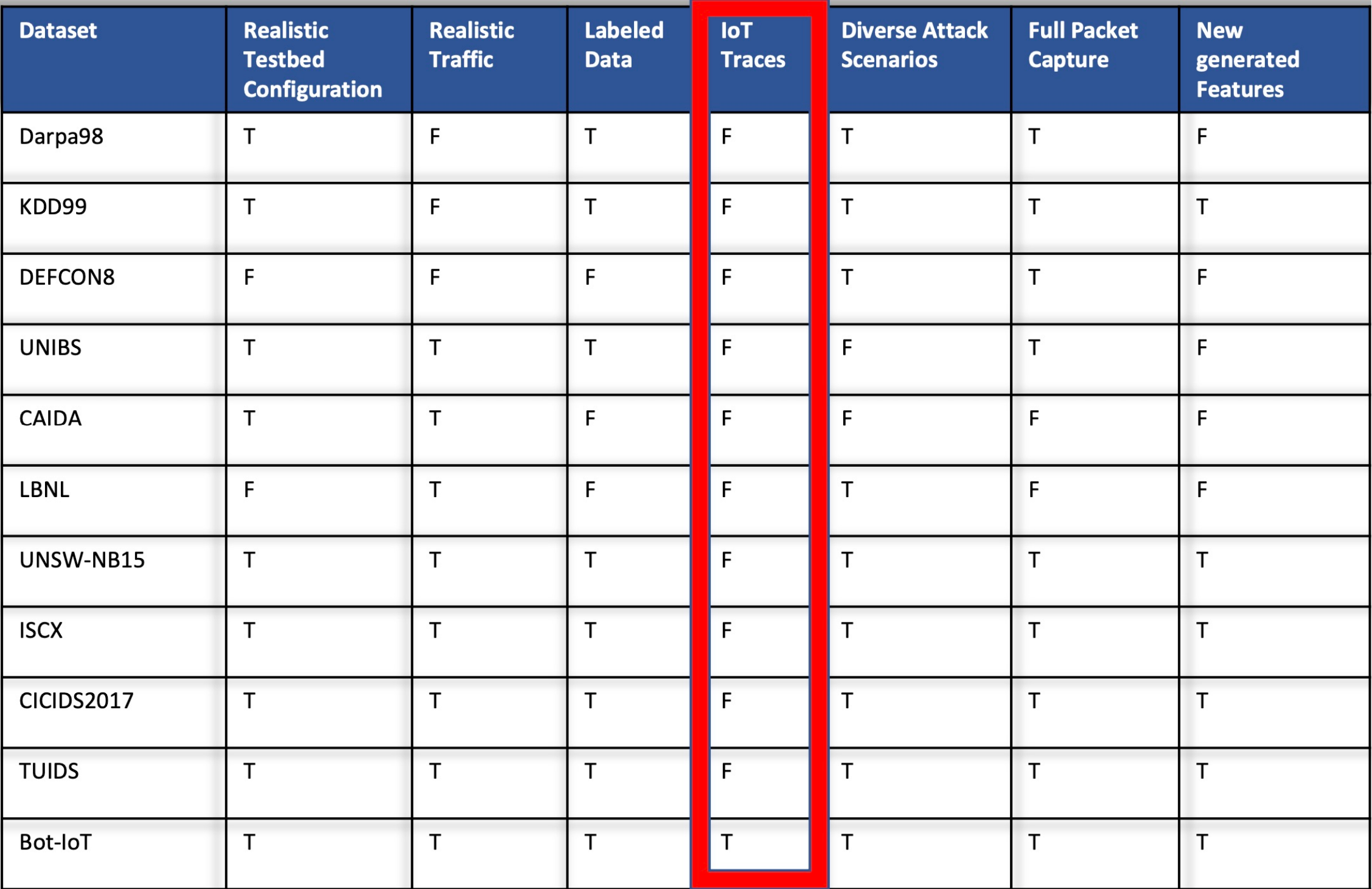}
\centering
\caption{Comparison of datasets used in intrusion detection (F=False, T=True).}
\label{fig:NIDS_datasets}
\end{figure}

There are various datasets related to intrusion detection~\cite{koroniotis2019towards, sharafaldinTowardGenerating2018, tavallaeeDetailedAnalysisKDD2009, moustafaUNSWNB15ComprehensiveData2015, shiraviDevelopingSystematicApproach2012, KDDCup1999Data}. These datasets are summarized in the Figure~\ref{fig:NIDS_datasets}. From this figure, we can see that only the Bot-IoT dataset contains traces of IoT network traffic. This makes it plausible for the analysis of intrusion detection in IoT networks.

The KDD dataset is a commonly used dataset for the evaluation of intrusion detection and contains seven weeks of network traffic including simulated attacks.
In \cite{tavallaeeDetailedAnalysisKDD2009}, Tavallaee et al.\ proposed NSL-KDD dataset, which is an extended version of the KDD dataset such that it does not contain redundant and duplicate records, the number of records is reasonable.
The UNSW-NB15 dataset is generated by IXIA PerfectStorm, Tcpdump, Argus, and Bro-IDS tools, which create some types of attacks~\cite{moustafaUNSWNB15ComprehensiveData2015}.
Shiravi et al.\ created ISCX dataset consisting of the seven days of both normal and abnormal activities~\cite{shiraviDevelopingSystematicApproach2012}.
Sharafaldin et al.\ proposed CICIDS2017 dataset, which contains common attacks resembling the real-world data, i.e., packet capture~\cite{sharafaldinTowardGenerating2018}.
In contrast to these studies, Koroniotis et al.\ focused on the IoT network and created the Bot-IoT dataset, which consists of the IoT network traffic including various types of attacks~\cite{koroniotis2019towards}.
We will describe the details of the Bot-IoT in Section~\ref{sec:Bot-IoT Dataset}.
Since we also focus on the NIDS in IoT networks, the Bot-IoT dataset is used in this thesis. In addition we also tackle the class imbalance problem by using the over-sampling method.

We note that our work is similar to a recent study conducted by~\cite{pokhrel2021iot}. However, the major differences are: (1) We removed all non-network features before any analysis is conducted, that is, \textit{pkSeqID}, \textit{seq}, \textit{dur}, \textit{mean}, \textit{stddev}, \textit{sum}, \textit{min}, and \textit{max} (2) We employ three algorithms to select the list of features needed for intrusion detection analysis (3) Whereas their study investigates three machine learning models namely  K-Nearest Neighbours (KNN), Naive Bayes and Multi-layer Perceptron, we investigated seven models, that is, Logistic  Regression,  Linear  SVC,  Linear  Kernel  SVM,  RBF Kernel SVM, Random Forest, Extreme Gradient Boosting, and Multi-layer Perceptron (4) We evaluate and select the best classifiers based on false positive rate, false negative rate, and inference time, on top of accuracy, recall, precision and F1-scores.

\newpage
\section{Background}
\label{sec:Background}
In this section, we introduce the Bot-IoT dataset and ML techniques used in this thesis.
\subsection{Bot-IoT Dataset}
\label{sec:Bot-IoT Dataset}
\begin{table}[t]
  \centering
  \caption{BoT-IoT dataset features.}
  \label{table:all_features_in_bot-iot_dataset}
  \scalebox{0.6}{
    \begin{tabular}{ll}
      \hline
      Feature& Description\\
      \hline
      \hline
      \textit{pkSeqID }&Row Identifier\\
      \textit{sbytes }& Source-to-destination byte count \\
      \textit{Stime }& Record start time \\
      \textit{dbytes}&Destination-to-source byte count \\
      \textit{flgs }& Flow state flags seen in trans- actions\\
      \textit{rate }&Total packets per second in transaction \\
      \textit{flgs number }& Numerical representation of feature flags\\
      \textit{srate }&Source-to-destination packets per second\\
      \textit{Proto }& Textual representation of transaction protocols present in network flow \\
      \textit{drate }&Destination-to-source packets per second \\
      \textit{proto\_number }& Numerical representation of feature proto \\
      \textit{TnBPSrcIP }&Total Number of bytes per source IP \\
      \textit{saddr }& Source IP address \\
      \textit{TnBPDstIP }&Total Number of bytes per Destination IP \\
      \textit{sport }& Source port number \\
      \textit{TnP\_PSrcIP }&Total Number of packets per source IP \\
      \textit{daddr }& Destination IP address \\
      \textit{TnP\_PDstIP }&Total Number of packets per Destination IP \\
      \textit{dport }& Destination port number\\
      \textit{TnP\_PerProto }&Total Number of packets per protocol\\
      \textit{pkts }& Total count of packets in transaction \\
      \textit{TnP\_Per Dport }&Total Number of packets per dport \\
      \textit{bytes }& Total number of bytes in transaction \\
      \textit{AR\_P\_Proto\_P\_SrcIP }&Average rate per protocol per Source IP. (calculated by pkts/dur) \\
      \textit{state }& Transaction state \\
      \textit{AR\_P\_Proto\_P\_DstIP }&Average rate per protocol per Destination IP \\
      \textit{state\_number }& Numerical representation of feature state\\
      \textit{ltime }& Record last time \\
      \textit{N\_IN\_Conn\_P\_SrcIP }&Number of inbound connections per source IP\\
      \textit{seq }& Argus sequence number \\
      \textit{N\_IN\_Conn\_P\_DstIP }&Number of inbound connections per destination IP \\
      \textit{dur }& Record total duration \\
      \textit{AR\_P\_Proto\_P\_Sport}&Average rate per protocol per sport \\
      \textit{mean }& Average duration of aggregated records \\
      \textit{AR\_P\_Proto\_P\_Dport }&Average rate per protocol per dport \\
      \textit{stddev }& Standard deviation of aggregated records \\
      \textit{Pkts\_P\_State\_P\_Protocol\_P\_DestIP }&Number of packets grouped by state of flows and protocols per destination IP\\
      \textit{sum }& Total duration of aggregated records \\
      \textit{Pkts\_P\_State\_P\_Protocol\_P\_SrcIP  }&Number of packets grouped by state of flows and protocols per source IP \\
      \textit{min }& Minimum duration of aggregated records \\
      \textit{attack}&Class label: 0 for Normal traffic, 1 for Attack Traffic \\
      \textit{max }& Maximum duration of aggregated records \\
      \textit{category }&Traffic category \\
      \textit{spkts }& Source-to-destination packet count \\
      \textit{subcategory }&Traffic subcategory \\
      \textit{dpkts }& Destination-to-source packet count\\
      \hline
    \end{tabular}
  }
\end{table}
\begin{table}[t]
  \centering
  \caption{Class distribution in the BoT-IoT dataset.}
  \label{table:Class distribution in the BoT-IoT dataset}
  \begin{tabular}{lr}
    \hline
    Class label & Number of samples\\
    \hline
    \hline
    Normal & 477\\
    Attack & 3,668,041\\
    \hline
  \end{tabular}
\end{table}

Koroniotis et al.\ developed a realistic testbed to simulate the IoT network traffic including various types of attacks, i.e., DDos, DoS, operating system and service scan, key logging and data theft attacks, and published the Bot-IoT dataset generated by this simulation~\cite{koroniotis2019towards}.
The Bot-IoT dataset contains more than 72 million records, each of which represents a network packet comprising of 43 features and is categorized into either a \textit{normal} packet or an \textit{attack} one.
Note that all the features in the dataset are shown in Table~\ref{table:all_features_in_bot-iot_dataset}.

To evaluate the reliability of this dataset, the authors proposed ML-based algorithms to detect attack packets and demonstrated good accuracy~\cite{koroniotis2019towards}.
However, the dataset used in this evaluation contains imbalanced normal and attack packets because the number of attack packets is much higher than that of normal ones, as shown in Table~\ref{table:Class distribution in the BoT-IoT dataset}.
This distribution is counterintuitive because we can usually monitor more normal packets than attack ones.
Such imbalanced data may cause performance degradation in the minority class.

In this thesis, we address the class imbalance problem in the Bot-IoT dataset to improve the performance by using the data sampling techniques.
\subsection{Machine Learning based Binary Classification}
\label{sec:Machine Learning based Binary Classification}
\subsubsection{Logistic Regression}
Logisitic regression~\cite{10.5555/1671238} is an extension to the linear regression to estimate a dependent variable, i.e., binary variable, from one or more independent variables, i.e, a feature vector $\bm{x}$, where $\bm{x} = (x_1, \ldots x_D)$ is a $D$ dimensional feature vector and $x_k$ ($1 \leq k \leq D$) is a feature value.

More precisely, given the independent variables $\bm{x}$, the logistic regression estimates the probability $p$ that the independent variables $\bm{x}$ belongs to the positive class.
The logistic regression uses the log-odds, i.e., logit function, where the logit function $z(\cdot)$ is defined as the logarithm of the odds ratio, 
\begin{align}
  z(p) = \ln\left(\frac{p}{1-p}\right).
  \label{eq:logit_function}
\end{align}
Here, we can interpret the logit function $z(\cdot)$ as the linear combination of independent variables, i.e, the dot product of a learnable weight vector $\bm{w} = (w_1, \ldots, w_D)$ and a feature vector $\bm{x}$.
Therefore, we can map the logit function $z(\cdot)$ to the value at the range of $[0, 1]$, i.e., the probability $p$, by using the inverse function of \eqref{eq:logit_function}.
\begin{align}
  p = \frac{1}{1-e^{-z}} = \frac{1}{1-e^{-\bm{w}^\mathsf{T}\cdot\bm{x}}}.
\end{align}
If the probability $p$ is higher than or equal to a certain threshold, i.e., 0.5, $\bm{x}$ belongs to the positive class.
The logistic regression aims at learning the weight vector $\bm{w}$ to maximize the logarithm of the likelihood function.
\subsubsection{Support Vector Machine}
A support vector machine (SVM)~\cite{bishopPattern2006} is one of the supervised learning models and aims at finding an appropriate separating hyperplane in high-dimensional feature spaces to discriminate between categories by using a kernel function, e.g., linear, polynomial, or radial basis function.
In case of binary classification, the SVM classifier calculates the hyperplane to distinguish between the data belonging to a certain class and the rest of data.
The appropriate hyperplane is obtained by maximizing the distance from the hyperplane to the closest point across both classes under the assumption where the training data are linearly separable. This maximum distance is called the \textit{maximum margin separator}. 

Consider a training dataset of $n$ points of the form
$\left(\mathbf{x}_{1}, y_{1}\right), \ldots,\left(\mathbf{x}_{n}, y_{n}\right)$
where the $y_{i}$ are either 1 or $-1$, each indicating the class to which the point ${\mathbf {x} _{i}}$ belongs. Each ${\mathbf {x} _{i}}$ is a real vector of dimensions $D$. Our aim is to find the hyperplane of maximum margin that divides the group of points ${\mathbf {x} _{i}}$ for which ${y_{i}=1}$, from the group of points for which ${y_{i}=-1}$. Our aim is formulated so that the distance between the hyperplane and the nearest point ${\mathbf {x} _{i}}$ from either group is maximized.
The hyperplane can be written as the set of points $\mathbf{x}$ which satisfy $\mathbf{w}^{T} \mathbf{x}-b=0\text {, }$ where $\mathbf {w}$  is the normal vector to the hyperplane. The parameter ${\tfrac {b}{\|\mathbf {w} \|}}$ determines the offset of the hyperplane from the origin along the normal vector  $\mathbf {w}$, that is, the margin. To compute the soft-margin SVM classifier is equivalent to minimizing the optimization problem:
\begin{align}
{\left[{\frac {1}{n}}\sum _{i=1}^{n}\max \left(0,1-y_{i}(\mathbf {w} ^{T}\mathbf {x} _{i}-b)\right)\right]+\lambda \|\mathbf {w} \|^{2}.} 
\end{align}
The \textit{primal problem} definition is obtained by reformulating the optimization problem above as:
\begin{align}
{\text{min}}{\frac {1}{n}}\sum _{i=1}^{n}\zeta _{i}+\lambda \|\mathbf {w} \|^{2}\text {, }
{\text{subject to }}y_{i}(\mathbf {w} ^{T}\mathbf {x} _{i}-b)\geq 1-\zeta _{i}\,{\text{ and }}\,\zeta _{i}\geq 0, \forall i.
\end{align}  \\
The variable ${\zeta _{i}=\max \left(0,1-y_{i}(\mathbf {w} ^{T}\mathbf {x} _{i}-b)\right)}$ is introduced for each ${ i\in \{1,\,\ldots ,\,n\}}.$ The \textit{primal problem} can be further simplified to give the dual maximization problem, that is the \textit{dual problem}. The \textit{dual problem} is a quadratic programming problem and it is defined below:
\begin{align}
{\text{max}\,f(c_{1}\ldots c_{n})=\sum _{i=1}^{n}c_{i}-{\frac {1}{2}}\sum _{i=1}^{n}\sum _{j=1}^{n}y_{i}c_{i}(\mathbf {x} _{i}^{T}\mathbf {x} _{j})y_{j}c_{j} \text {, }}
\end{align}
${\text{subject to }}\sum _{i=1}^{n}c_{i}y_{i}=0,{\text{and }}0\leq c_{i}\leq {\frac {1}{2n\lambda }}, \forall i.$\\
The variables $c_{i}$ are defined such that ${\mathbf {w} =\sum _{i=1}^{n}c_{i}y_{i}\mathbf {x} _{i}}$. Moreover, ${c_{i}=0}$ when ${\mathbf {x} _{i}}$ lies on the correct side of the margin, and ${ 0<c_{i}<(2n\lambda )^{-1}}$ when ${\mathbf {x} _{i}}$ lies on the margin's boundary. It follows that ${\mathbf {w}}$ can be written as a linear combination of the support vectors. The offset, $b$ can be recovered by finding an ${\mathbf {x} _{i}}$ on the margin's boundary and solving ${b=\mathbf {w} ^{T}\mathbf {x} _{i}-y_{i}.}$

The sequential minimal optimization (SMO) algorithm was proposed, which is a fast algorithm for solving the dual maximization problem in order to train SVMs~\cite{platt1998sequential}.

\subsubsection{Random Forest}
Random forest (RF)~\cite{10.1023/A:1010933404324} is a supervised learning method and ensemble learning using multiple decision trees, where the ensemble learning algorithm combines multiple ML algorithms to obtain higher performance.
The RF creates multiple decision trees, each of which is generated from a different training subset provided by the training dataset sampled with replacement.
Each decision tree is independently trained and outputs a classification result.
The RF outputs a classification result based on majority voting after aggregating all the classification results of all the decision trees.

\subsubsection{Extreme Gradient Boosting}
The extreme gradient boosting (XGBoost) algorithm~\cite{chenXGBoostScalableTree2016} is an extended version of the gradient boosting decision tree (GBDT)~\cite{yeStochastic2009} in a distributed manner.
The GBDT is ensemble learning based on multiple decision trees similar to the RF.
The GBDT, however, has a different aspect from the RF in terms of the ensemble algorithm.

The ensemble algorithm used in the RF combines full decision trees in a parallel manner while that used in the GBDT produces a classifier by combining weaker decision trees in a sequential manner.

\begin{figure}[t]
\includegraphics[width=0.8\columnwidth]{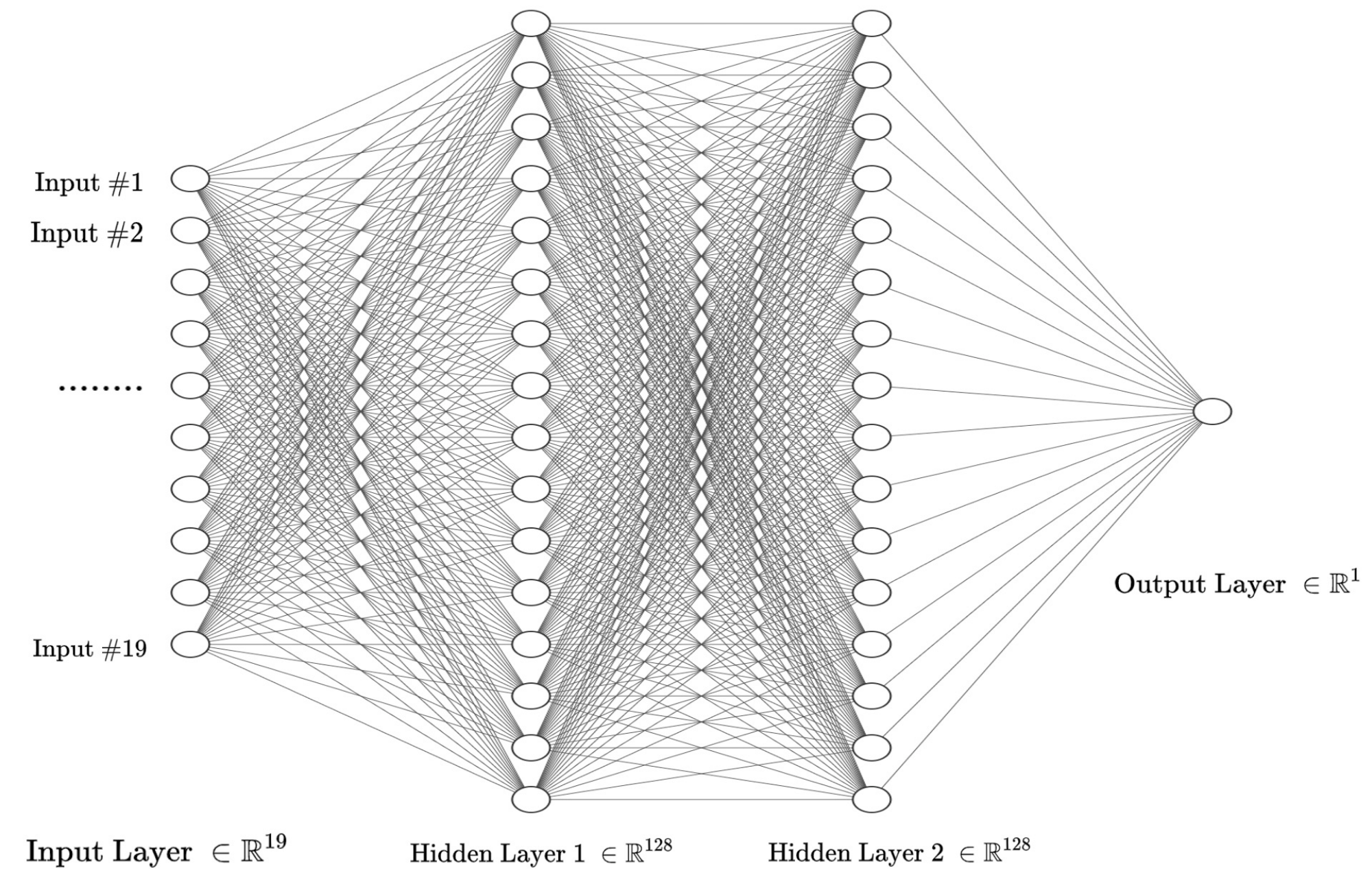}
\centering
\caption{An example of MLP with two hidden layers.}
\label{fig:mlp}
\end{figure}
\subsubsection{Multi-layer Perceptron Neural Network}
Multi-layer perceptron (MLP) is a multiple feedforward artificial neural network with one or more hidden layers between the input and output layers and succeeds in many applications including classification.
Fig.~\ref{fig:mlp} illustrates an example of MLP with two hidden layers.
The MLP can be defined as a directed graph with multiple node layers.
The left (resp.\ right) side layer indicates the input (resp.\ output) layer and the others mean the hidden layers.
The MLP is a fully connected network where every node on a certain layer has connections with a certain weight to all the nodes on the next layer, except for the output layer.
Each node corresponds to a processing unit with a nonlinear activation function, e.g., rectified linear unit function, except for the nodes on the input layer.
Thanks to nonlinear mapping, the MLP can approximate to any continuous function, which is well-known as universal approximation theorem~\cite{cybenkoApproximationSuperpositionsSigmoidal1989}.
A backpropagation algorithm~\cite{rojas2013neural}, which is one of the supervised learning methods, is adopted to train MLP.
\subsection{Data Sampling Methods}
\label{sec:Data Sampling Methods}
The class imbalance problem arises from the real-world data.

In general, such imbalanced data causes performance degradation in the minority class.
To tackle this issue, some sampling techniques were proposed~\cite{leevySurveyAddressingHighClass2018, Chawla_2002, hanBorderlineSMOTENewOverSampling2005, heADASYNAdaptiveSynthetic2008, bunkhumpornpatSafeLevelSMOTESafeLevelSyntheticMinority2009}.
Random minority over-sampling (ROS) and random majority under-sampling (RUS) are commonly used and increase the sensitivity of a classifier to the minority class.
The ROS randomly duplicates samples with the minority class while the RUS randomly discards samples with the majority class from the dataset.
Chawla et al.\ proposed an over-sampling method called synthetic minority over-sampling technique (SMOTE), where the minority class is over-sampled by creating synthetic data~\cite{Chawla_2002}.

The SMOTE algorithm aims at generating synthetic samples with the minority class according to a certain sampling rate $N$ based on the imbalanced proportion by the following procedures.
This algorithm first finds the $k$-nearest neighbors $\X_k(\bm{x}_i) = \{\bm{x}_1 \ldots, \bm{x}_k\}$ of each sample $\bm{x}_i \in \X$ by calculating the Euclidean distance between $\bm{x}_i$ and every sample $\bm{x}_j \in \X$ $(i \neq j)$, and selecting samples with the smallest Euclidean distance from the original sample.
Here, we define $\X$ $(X=|\X|)$ as a set of samples with the minority class and $\X_k(\bm{x}_i)$ as $k$-nearest neighbors of sample $\bm{x}_i$.
Next, for each sample $\bm{x}_i \in \X$, the algorithm randomly selects one of the $k$-nearest neighbors $\X_k(\bm{x}_i)$, i.e., $\bm{x}_l$ $(1 \leq l \leq k)$, and generates a new synthetic sample $\bm{x}'$ as follows:
\begin{align}
  \bm{x}' = \bm{x}_i + \gamma(\bm{x}_l - \bm{x}_i),
  \label{eq:smote}
\end{align}
where $\bm{\gamma}$ denotes a vector of uniform random values between 0 and 1.
We repeat the above selection and generation until $N$ synthetic samples are generated.
Applying these procedures to all samples in $\X$, we consequently obtain $NX$ synthetic samples with minority class.

In addition to the SMOTE algorithm, several variants of the SMOTE algorithm have been proposed~\cite{hanBorderlineSMOTENewOverSampling2005, heADASYNAdaptiveSynthetic2008, bunkhumpornpatSafeLevelSMOTESafeLevelSyntheticMinority2009}.
We should note that such over-sampling techniques may change the nature of the dataset but improve the performance of the classifiers~\cite{vanhulseExperimentalPerspectivesLearning2007}.
\newpage
\section{Proposed Method}
\label{sec:Proposed Method}
In this section, we propose binary classification to deal with the class imbalance problem in the Bot-IoT dataset.
\begin{figure*}[h]
  \centering
    \includegraphics[width=0.99\columnwidth]{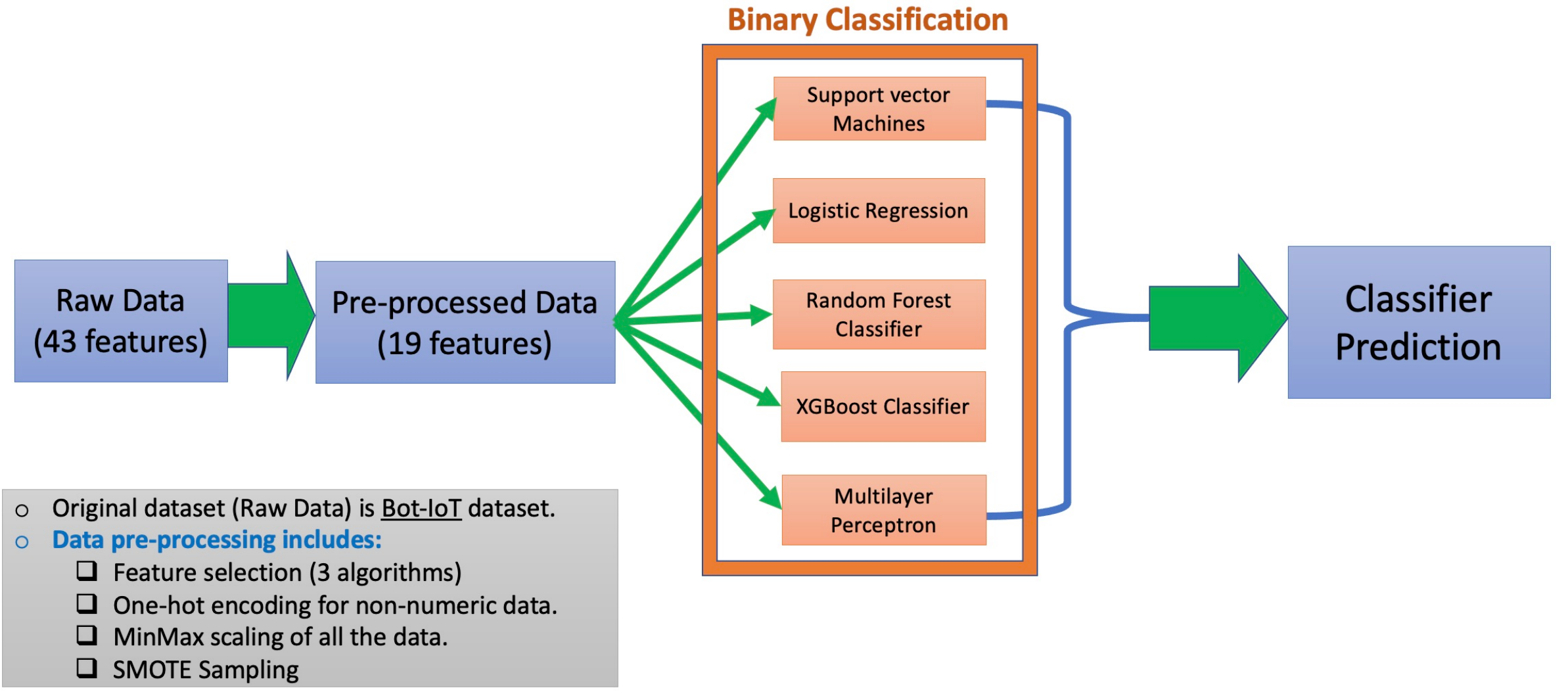}
    \caption{Intrusion Detection analysis based on Bot-IoT dataset.}%
    \label{fig:Methodology}
\end{figure*}

\subsection{Overview}
\label{sec:Overview}
As mentioned in Section~\ref{sec:Bot-IoT Dataset}, the Bot-IoT dataset contains imbalanced normal and attack packets because the number of attack packets is much higher than that of normal ones.
This may cause performance degradation in the minority class.
We propose the binary classification method to address the class imbalance problem in the Bot-IoT dataset.
The proposed method consists of mainly three stages, i.e., preprocessing, data sampling, and binary classification.
We first conduct the preprocessing to tackle the curse of dimensionality, which will be shown in Section~\ref{sec:Preprocessing}.
To investigate the positive (resp.\ negative) impact of the balanced (resp.\ imbalanced) dataset, we use the SMOTE algorithm, which generates the synthetic samples such that number of normal packets is equivalent to that of attack ones.
(See the details of the data sampling in Section~\ref{sec:Data Sampling}.)
The proposed binary classifiers aim at detecting the attack packets in Bot-IoT dataset.
As for performance comparison purposes, we adopt several binary classifiers, i.e., logistic regression, SVM, RF, XGBoost, and MLP.
We will describe the details of the proposed binary classifiers in Section~\ref{sec:Binary Classifiers}.
\subsection{Preprocessing}
\label{sec:Preprocessing}
\begin{figure}[t]%
  \centering
  \includegraphics[width=0.99\columnwidth]{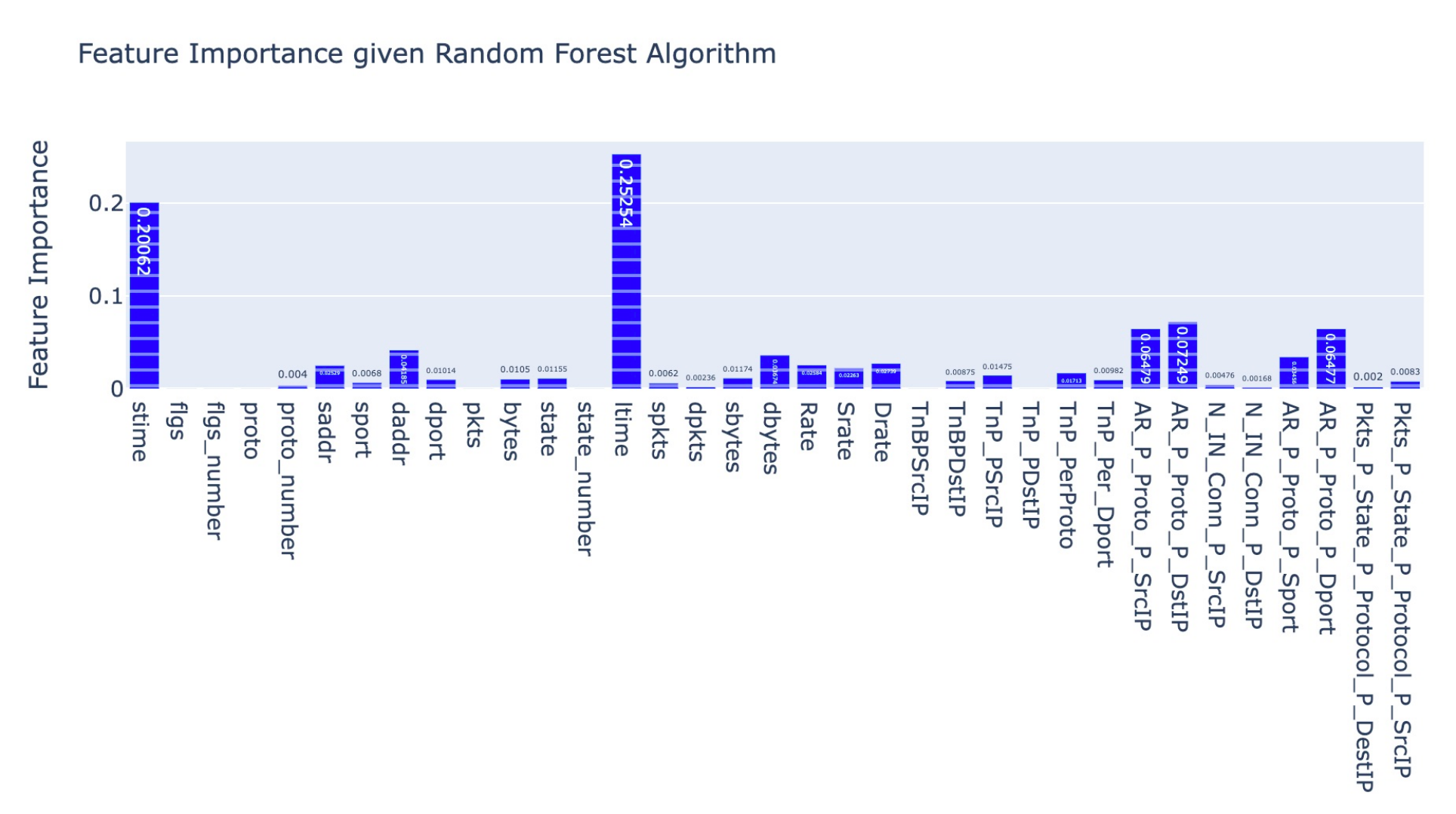}
  \caption{Feature importance with the random forest algorithm.}%
  \label{fig:feature_importance_with_random_forest_algorithm}
\end{figure}
\begin{figure}[t]%
  \centering
    \includegraphics[width=0.99\columnwidth]{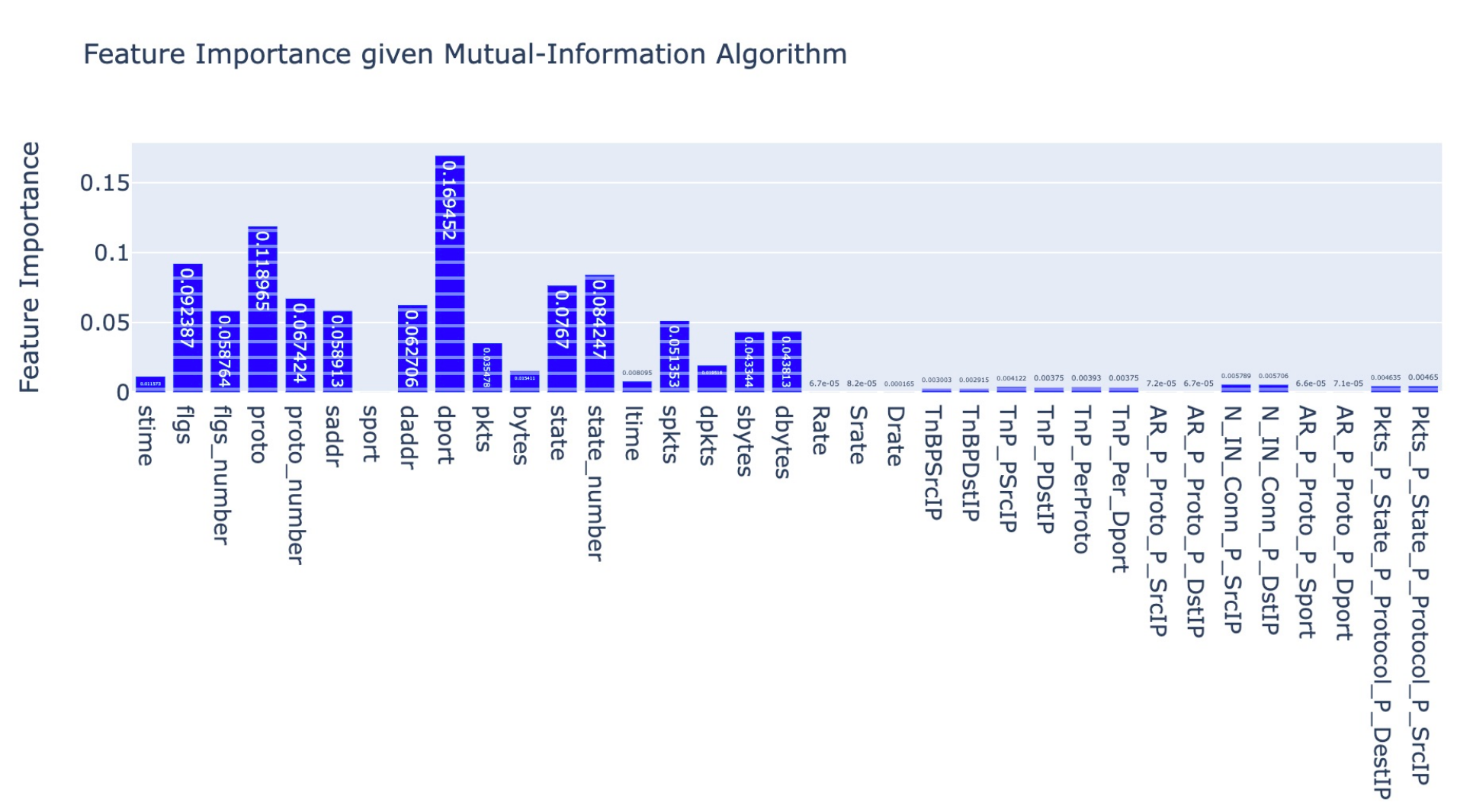}
    \caption{Feature importance with the mutual information algorithm.}%
    \label{fig:feature_importance_with_mutual_information_algorithm}
\end{figure}
\begin{figure}[t]%
  \centering
    \includegraphics[width=0.99\columnwidth]{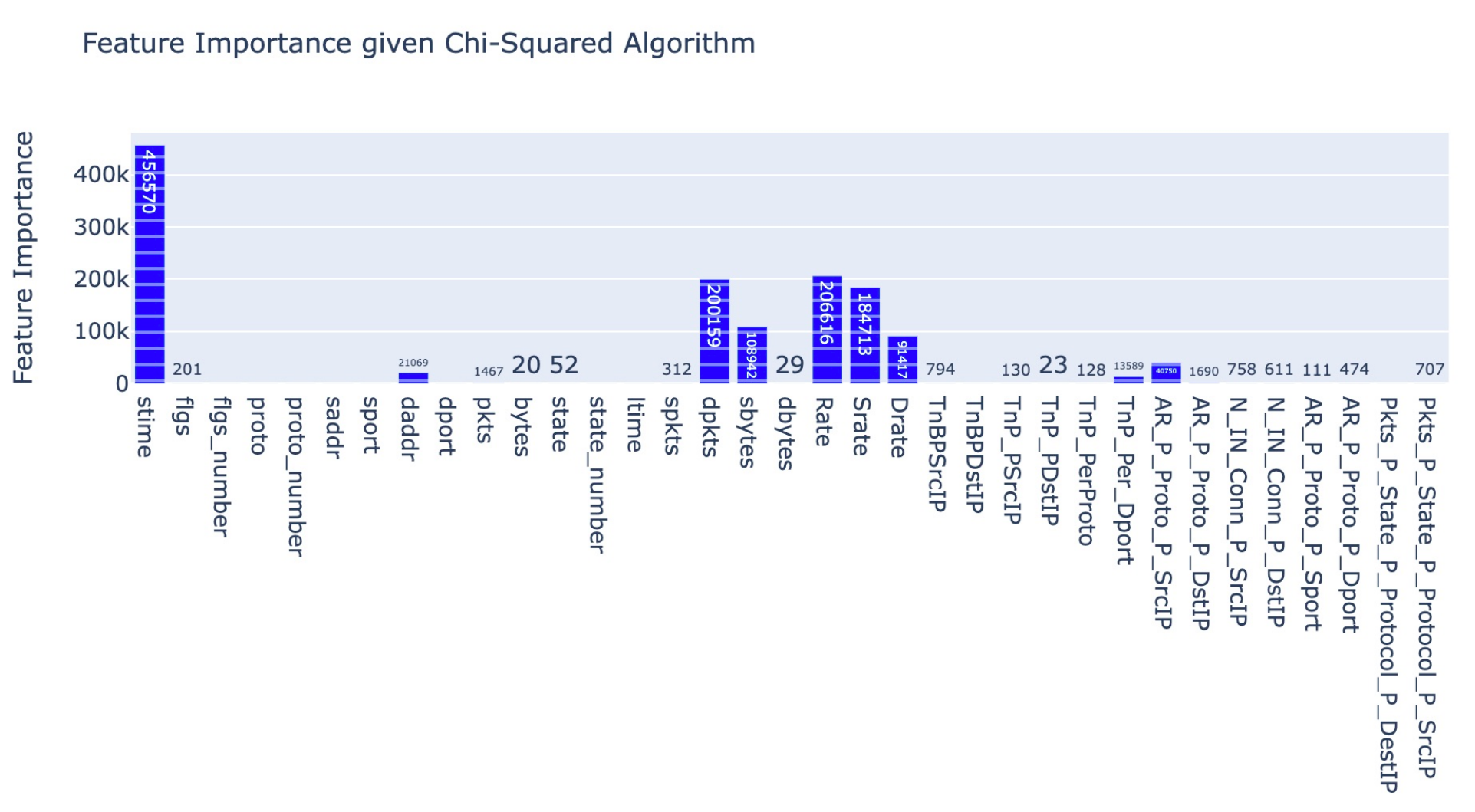}
    \caption{Feature importance with the chi-squared algorithm.}%
    \label{fig:feature_importance_with_chi-squared_algorithm}
\end{figure}
\begin{table}[t]
  \caption{Features selected by each feature selection algorithm.}
  \centering
  \label{table:feature_selection}
  \begin{tabular}{lll}
    \hline
    Chi-squared & Mutual information & Random forest\\
    \hline
    \hline
    \textit{srate}                  & \textit{dport}         & \textit{ltime} \\
    \textit{sport}                  & \textit{proto}         & \textit{stime} \\
    \textit{AR\_P\_Proto\_P\_Sport} & \textit{flgs}          & \textit{AR\_P\_Proto\_P\_DstIP}\\
    \textit{AR\_P\_Proto\_P\_SrcIP} & \textit{state}         & \textit{AR\_P\_Proto\_P\_SrcIP}\\
    \textit{AR\_P\_Proto\_P\_DstIP} & \textit{proto\_number} & \textit{AR\_P\_Proto\_P\_Dport}\\
    \textit{rate}                   & \textit{daddr}         & \textit{daddr}\\
    \textit{ARP\_Proto\_P\_Dport}   & \textit{saddr}         & \textit{AR\_P\_Proto\_P\_Sport}\\
                                    & \textit{flgs\_number}  & \textit{rate}\\
                                    &                        & \textit{TnP\_Per\_Proto}\\
                                    &                        & \textit{bytes}\\
    \hline
  \end{tabular}
\end{table}

We use the Bot-IoT dataset~\cite{koroniotis2019towards} containing 43 features.
Recall that all the features in the Bot-IoT dataset are shown in Table~\ref{table:all_features_in_bot-iot_dataset}.
To tackle curse of dimensionality, we select important features related to the prediction accuracy from the original features by using feature selection algorithms based on three metrics, i.e., random forest~\cite{breiman2001random}, mutual information~\cite{rossMutualInformationDiscrete2014}, and chi-squared algorithm~\cite{doi:10.1080/14786440009463897}.
Before executing the feature selection algorithms, we remove the features not related to the typical characteristics of the network traffic, i.e., \textit{pkSeqID}, \textit{seq}, \textit{dur}, \textit{mean}, \textit{stddev}, \textit{sum}, \textit{min}, and \textit{max}, from the original Bot-IoT dataset.
Appying each algorithm to the Bot-IoT dataset with the remaining features, we calculate the feature importance scores with the three algorithms, as shown in Figs.~\ref{fig:feature_importance_with_random_forest_algorithm} through \ref{fig:feature_importance_with_chi-squared_algorithm}.
Figs.~\ref{fig:feature_importance_with_random_forest_algorithm} through \ref{fig:feature_importance_with_chi-squared_algorithm} illustrate the feature importance by applying each feature selection algorithm, respectively.
We manually select the important features such that feature importance derived by each algorithm is higher than a certain threshold.
Table~\ref{table:feature_selection} shows the features selected by each feature selection algorithm.
We use the 19 features, which is the union set of the features selected by the three feature selection algorithms.

Since this dataset contains categorical features, we apply the one-hot encoding to the categorical features, which results in the numeric features.
We also apply the min-max normalization to each feature to improve the classifier performance.
More precisely, the min-max normalization translates the $d$th ($1 \leq d \leq D$) feature value $x_{i,d}$ of the $i$th sample ($D$ dimensional feature vector) $\bm{x}_i=(x_{i,1},\ldots,x_{i,D})$ to the following value $x_{i,d}^{\mathrm{normalized}}$:
\begin{align}
  x_{i,d}^{\mathrm{normalized}} = \frac{x_{i,d} - \min_{\bm{x}_j \in \X} x_{j,d}} {\max_{\bm{x}_j \in \X} x_{j,d} - \min_{\bm{x}_j \in \X} x_{j,d}},
\end{align}
where $\X$ denotes a set of all samples.
As a result, we obtain the normalized feature value between zero and one.
\subsection{Data Sampling}
\label{sec:Data Sampling}
Since the Bot-IoT dataset contains imbalanced normal and attack packets, the classifier performance will be skewed.
To address the class imbalance problem, we apply the SMOTE algorithm~\cite{Chawla_2002} to the dataset with the minority class, i.e., normal packets, which generates the synthetic samples from the dataset with the minority class such that the number of normal packets is equivalent to that of attack ones.
Remember that the detail of the SMOTE algorithm is described in Section~\ref{sec:Data Sampling Methods}.
We use the oversampled dataset only for the training phase to mitigate the skewed classifier performance.
\subsection{Binary Classifiers}
\label{sec:Binary Classifiers}
The proposed binary classifier aims at detecting the attack packets from the packets in the Bot-IoT dataset.
As for performance comparison purposes, we use several binary classifiers, i.e., logistic regression, linear SVM Classification (LinearSVC), linear kernel SVM, radial basis function (RBF) kernel SVM, random forest, XGBoost, and MLP.
To train the SVM model, we adopt the two types of the solvers, i.e., \textit{liblinear} solver~\cite{10.5555/1390681.1442794} and \textit{libsvm} one~\cite{10.1145/1961189.1961199}.
The liblinear solver is used for the Linear SVC without kernel transform while the libsvm sovler is used for the Linear and RBF Kernel SVMs.
As a result, the former can quickly solve the SMO algorithm and deal with a large amount of data, compared with the latter.

To realize the malicious packet detection, each of them is trained by using the preprocessed training dataset after applying the SMOTE algorithm.
Given the preprocessed packet as the input, the proposed binary classifier involves categorizing the packet within a specific class, i.e., a normal packet or an attack one.

\newpage
\section{Numerical Results}
\label{sec:Numerical Results}
In order to verify the positive (resp.\ negative) impact of the balanced (resp.\ imbalanced) dataset on the classifier performance, we evaluate the proposed binary classifiers.
We first explain the evaluation settings in Section~\ref{sec:Evaluation Settings}.
We show the fundamental characteristics of the proposed binary classifiers in Section~\ref{sec:Performance Comparison among Classifiers}.
Finally, we further demonstrate the impact of data sampling on the classifier performance in Section~\ref{sec:Impact of Data Sampling}.
\subsection{Evaluation Settings}
\label{sec:Evaluation Settings}
\begin{table}[t]
  \centering
  \caption{Training and testing dataset sizes.}
  \label{table:Size of the BoT-IoT dataset}
  \scriptsize
  \begin{tabular}{lrrrr}
  \hline
    Scenario & \multicolumn{2}{c}{Training dataset size} & \multicolumn{2}{c}{Testing dataset size}\\
    & attack packet & normal packet & attack packet & normal packet\\
  \hline
  \hline
    Imbalanced data scenario & 2,457,583 & 324 & 1,210,458 & 153\\
    Balanced data scenario   & 2,457,583 & 2,457,583 & 1,210,458 & 153\\
  \hline
  \end{tabular}
\end{table}

\begin{table}[t]
  \caption{Confusion matrix.}
  \centering
  \label{table:Confusion_matrix}
  \scriptsize
  \begin{tabular}{ll|ll}
    \hline
    & & \multicolumn{2}{c}{Prediction class}\\
    & & Attack packet & Normal packet\\
    \hline
    \hline
    \multirow{2}{*}{Actual class} & Attack packet & True Positive (TP) & False Negative (FN)\\
                                  & Normal packet & False Positive (FP) & True Negative (TN)\\
    \hline
  \end{tabular}
\end{table}
We use the server with Intel CPU Xeon Gold 6226R 16~core and 200~GB memory and with NVIDIA GeForce RTX 3090 GPU and 25.45~GB memory running on CUDA Version 11.1.
We implement the proposed binary classifiers, i.e., logistic regression, LinearSVC, linear kernel SVM, radial basis function (RBF) kernel SVM, random forest, XGBoost, and MLP, by using the python libraries, i.e., scikit-learn~\cite{pedregosa2011scikit} and Pytorch~\cite{NEURIPS2019_9015}.

As for evaluation, we use the hold-out method~\cite{yadavAnalysisKFoldCrossValidation2016} to evaluate each of the proposed classifiers where the 67\% of dataset is used for training while the rest of dataset is used for testing.
Table~\ref{table:Size of the BoT-IoT dataset} presents the training and testing dataset sizes in case of the imbalanced and balanced data scenarios, respectively.
To investigate impact of class imbalance, we prepare the two types of the trained models, i.e., the model trained on imbalanced data and that trained on balanced data.
In the former case, we use the dataset without the SMOTE algorithm for training.
On the other hand, we use the dataset with the SMOTE algorithm for training in the latter case.
Note that we apply the SMOTE algorithm to only the training dataset.
As for the SMOTE algorithm parameter, we adopt the number $k=5$ of nearest neighbors to construct the synthetic samples.

There are possible four cases for the binary classification result of each packet, i.e, true positive (TP), true negative (TN), false negative (FN), and false positive (FP).
Table~\ref{table:Confusion_matrix} presents a confusion matrix used in this thesis.
TP (resp.\ TN) indicates the case where the binary classifier correctly predicts the attack (resp.\ normal) packet.
On the other hand, FN (resp.\ FP) refers to the case where the packet is within the attack (resp.\ normal) class but the binary classifier performs incorrect prediction for the attack (resp.\ normal) packet.

As for evaluation metrics, we use \textit{accuracy}, \textit{recall}, \textit{precision}, \textit{false negative rate} (FNR), \textit{fales positive rate} (FPR), and \textit{F1-score}.
These metrics can be calculated as follows:
\begin{align}
  \mathrm{accuracy}&=\frac{TP+TN}{TP+FN+FP+TN},
  \\
  \mathrm{recall}&=\frac{TP}{TP+FN},
  \\
  \mathrm{precision}&=\frac{TP}{TP+FP},
  \\
  \mathrm{FNR}&=\frac{FN}{TP+FN},
  \\
  \mathrm{FPR}&=\frac{FP}{FP+TN},
  \\
  \text{F1-score}&=2 \cdot \frac{\mathrm{precision} \cdot \mathrm{recall}}{\mathrm{precision}+\mathrm{recall}},
\end{align}
where $TP$, $TN$, $FN$, and $FP$ represent the numbers of TP, TN, FN, and FP events, respectively.
We also use the area under a receiver operating characteristic (ROC) curve, i.e., \textit{AUC-score}, where an ROC curve is a curve plotting the TP rate against the FP rate at different classification thresholds and AUC-score indicates the entire two-dimensional area under the entire ROC curve.
In addition, we use \textit{inference time}, which indicates the actual time required for inference on all the testing data.
\subsection{Performance Comparison among Classifiers}
\label{sec:Performance Comparison among Classifiers}
\begin{table}[t]
  \centering
  \caption{Performance comparison among the proposed classifiers trained on imbalanced data.}
  \label{table:performance_comparison_imbalanced_scenario}
  \tiny   
  \begin{tabular}{lrrrrrrrr}
  \hline
    Scheme & accuracy & recall & precision & FNR & FPR & F1-score & AUC-score & inference time~$[\mathrm{s}]$\\
  \hline
  \hline
    Logistic regression & 99.9960 & 99.9999 & 99.9961 & 0.00008 & 30.72 & 99.9961 & 84.6404 & 0.05\\
    Linear SVC          & 99.9964 & 99.9995 & 99.9968 &  0.0004 & 24.84 & 99.9968 & 87.5814 & 0.05\\
    Linear kernel SVM   & 99.9960 & 100.0 & 99.9960 & 0.000 & 31.373 & 99.9960 & 84.3137 & 9.72\\
    RBF kernel SVM      & 99.9985 & 99.9999 & 99.9986 & 0.00008 & 10.457 & 99.9986 & 94.7712 & 27.06\\
    Random forest       & 99.8018 & 99.8032 & 99.9985 &  0.196 & 11.764 & 99.9985 & 94.0192 & 4.37\\
    XGBoost             & 99.9950 & 99.9976 & 99.9974 &  0.0023 & 20.26 & 99.9974 & 89.8680 & 0.43\\
    MLP                 & 99.8619 & 99.8697 & 99.9922 & 0.130 & 53.10 & 99.9922 & 73.3811 & 0.0012\\
  \hline
  \end{tabular}
\end{table}
In this section, we focus on the fundamental characteristics of the proposed binary classifiers.
Table~\ref{table:performance_comparison_imbalanced_scenario} presents the performance comparison among the proposed binary classifiers trained on the imbalanced dataset.
We first observe that all the proposed classifiers achieve high accuracy yet the training dataset contains class imbalance.
Focusing on recall, we confirm that all the proposed classifiers achieve over 99\% recall.
This indicates the proposed classifiers can correctly detect the attack packets.
In terms of precision, the proposed classifiers also achieve the high performance.
Next, focusing on the FPR, we observe that all the proposed classifiers have relatively high FPR, compared with FNR.
This is because the performance of the classifier is skewed towards the ``attack'' class due to the imbalanced training dataset, which results in a large number of false positive events.
From the viewpoint of AUC, the RBF kernel SVM and Random forest exhibit the high performance, i.e., 94.01\% and 94.07\%, while the AUC-score of the MLP is 73.38\%.

Next, we focus on the inference time.
We observe that logistic regression, Linear SVC, XGBoost, and MLP exhibit small inference time less than one second while Linear and RBF kernel SVMs and Random forest show much higher inference time than them.
The random forest requires long inference time to average the classification results obtained from the 1000 decision trees used for the binary classification.
Since Linear and RBF kernel SVM adjust the hyperplane such that the samples are distinguished into their class, they require the higher inference time.

\subsection{Impact of Data Sampling}
\label{sec:Impact of Data Sampling}
\begin{table}[t]
  \centering
  \caption{Performance comparison among the proposed classifiers trained on the balanced data.}
  \label{table:performance_comparison_balanced_scenario}
  \tiny
  \begin{tabular}{lrrrrrrrr}
  \hline
    Scheme & accuracy & recall & precision & FNR & FPR & F1-score & AUC-score & inference time~$[\mathrm{s}]$\\
    \hline
  \hline
    Logistic regression & 99.9722 & 99.9724 & 99.9998 & 0.027 & 1.307 & 99.9998 & 99.3326 & 0.05\\
    Linear SVC          & 99.9786 & 99.9788 & 99.9998 & 0.021 & 1.307 & 99.9998 & 99.3358 & 0.05\\
    Linear kernel SVM   & 99.6527 & 99.6535 & 99.9991 & 0.346 & 6.535 & 99.9991 & 96.5587 & 44.10\\
    RBF kernel SVM      & 99.9982 & 99.9984 & 99.9998 & 0.00156 & 1.307 & 99.9998 & 99.3456 & 76.66\\
    Random forest       & 99.9956 & 99.9957 & 99.9998 & 0.0042 & 1.307 &  99.9998 & 99.3442 & 4.35\\
    XGBoost             & 99.9945 & 99.9947 & 99.9998 & 0.0052& 1.307 & 99.9998 & 99.3437 & 0.45\\
    MLP                 & 99.9761 & 99.9781 & 99.9819 & 0.021 & 0.027 & 99.9819 & 99.9755 & 0.0012\\
  \hline
  \end{tabular}
\end{table}
In this section, we focus on the impact of class imbalance on the classifier performance.
Table~\ref{table:performance_comparison_balanced_scenario} presents the performance comparison among the proposed binary classifiers trained in the balanced dataset.
Comparing Table~\ref{table:performance_comparison_imbalanced_scenario}, we first confirm that all the proposed classifiers trained on the balanced dataset can exhibit almost the same accuracy, recall, and precision compared with those trained on the imbalanced dataset, thanks to the oversampling method.
We observe that the FPR is drastically improved because the proposed classifier is trained by using the large amount of data within the normal class.

This result shows that the normal packet cannot be blocked from reaching the IoT devices by introducing the balanced dataset.

Next, we focus on how the oversampling method affects the inference time.
We observe that the performance of Linear and RBF kernel SVM trained on the balanced data degrades in terms of inference time compared with those trained on the imbalanced data.
This is because these classifiers require time to fit the large amount of samples by finding the hyperplane.
On the other hand, the rest of the proposed classifiers trained in the balanced data exhibit almost the same inference time as those trained in the imbalanced data.

\begin{figure*}
    \centering
            \begin{subfigure}[!htb]{0.8\textwidth}
                \centering
                \includegraphics[width=\textwidth]{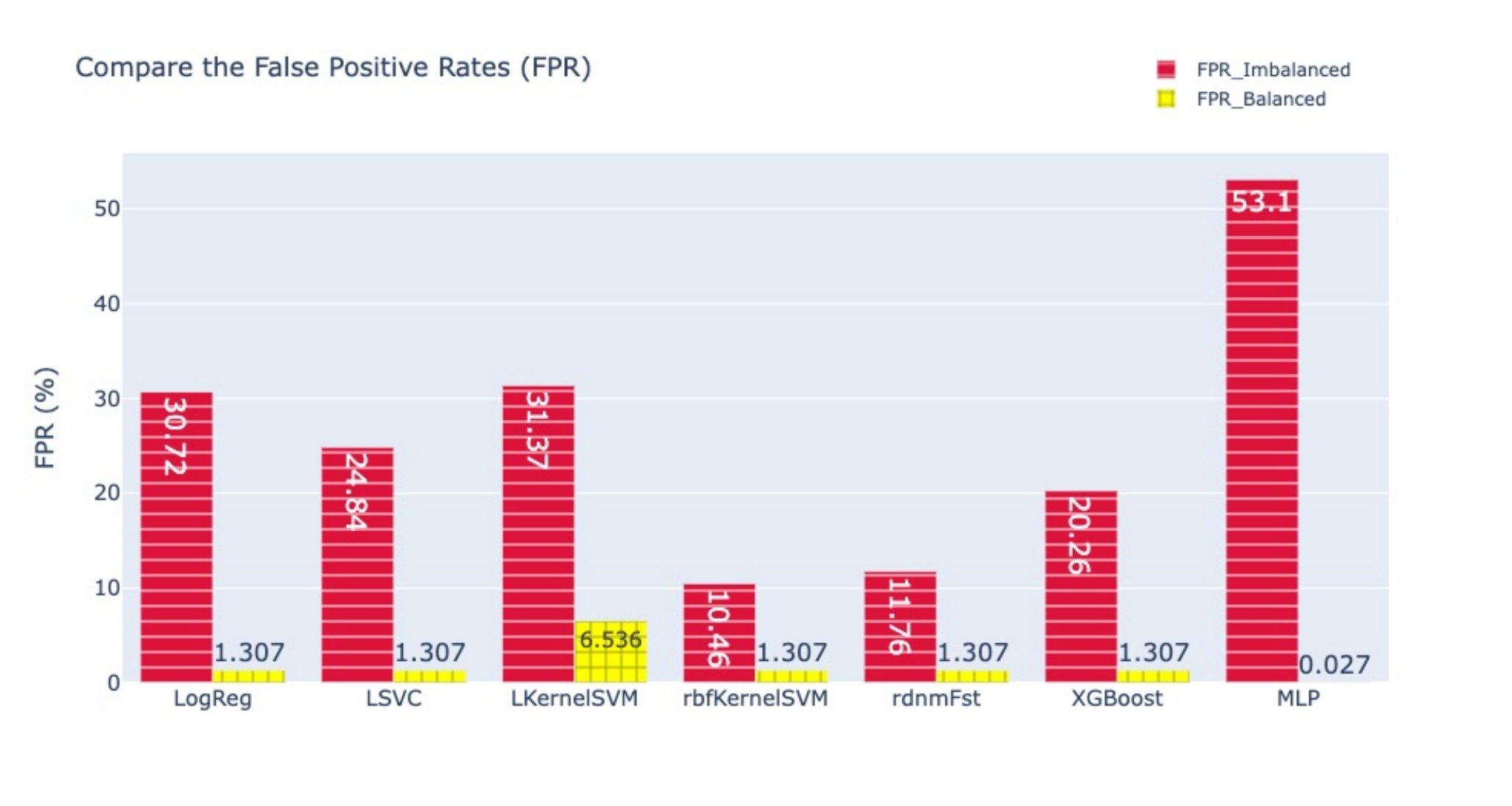}
                \subcaption{FPR on Imbalanced and Balanced data.}
                \label{fig:FPR}
            \end{subfigure}%
            \hfill
            \begin{subfigure}[!htb]{0.8\textwidth}
                \centering
                    \includegraphics[width=\textwidth]{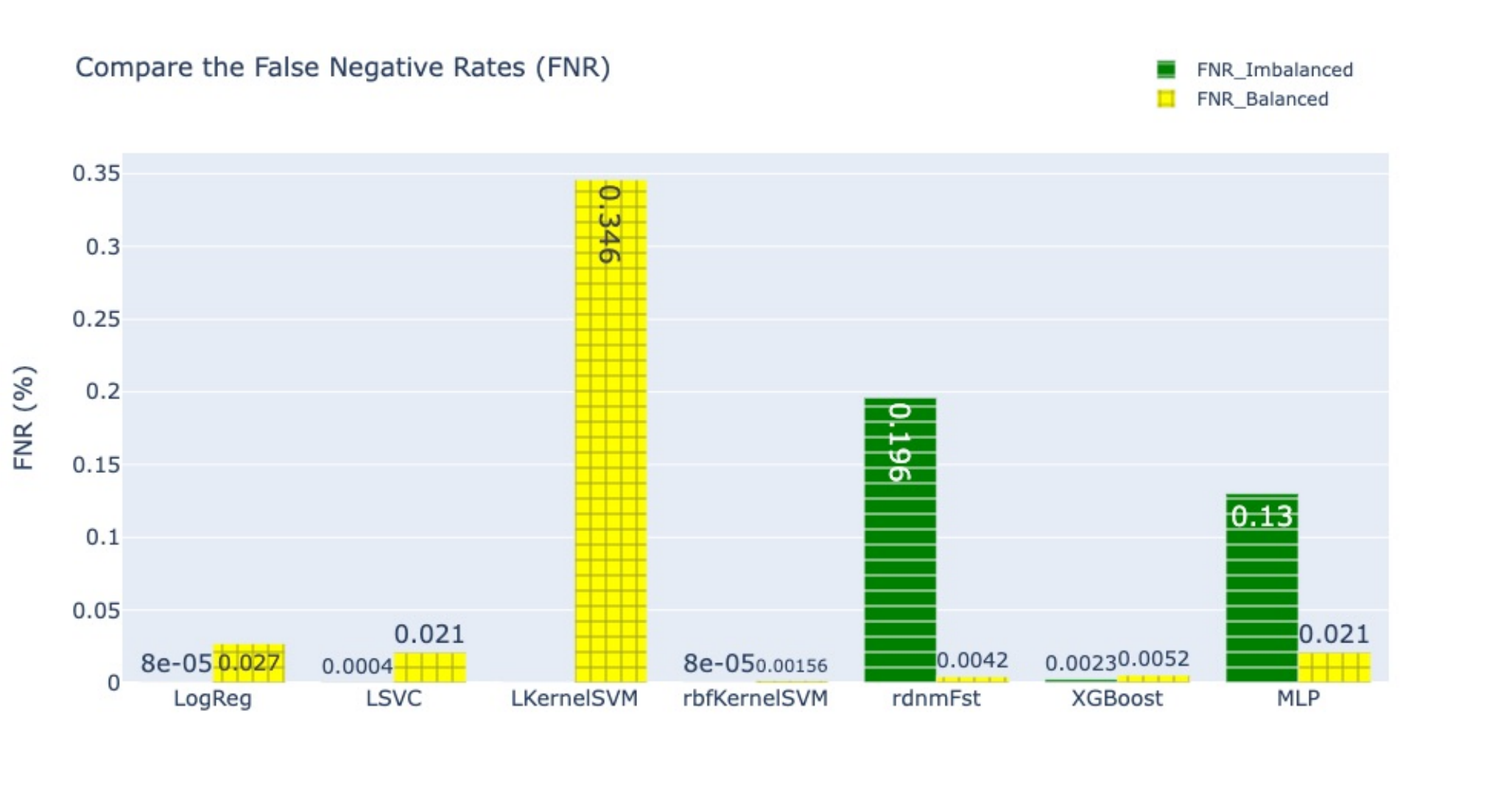}
                    \subcaption{FNR on Imbalanced and Balanced data.}
                    \label{fig:FNR}
            \end{subfigure}%
    \caption{FPR and FNR on Imbalanced and Balanced data.}
\end{figure*}

\begin{figure*}
 \centering
        \begin{subfigure}[!htb]{0.8\textwidth}
            \centering
                \includegraphics[width=\textwidth]{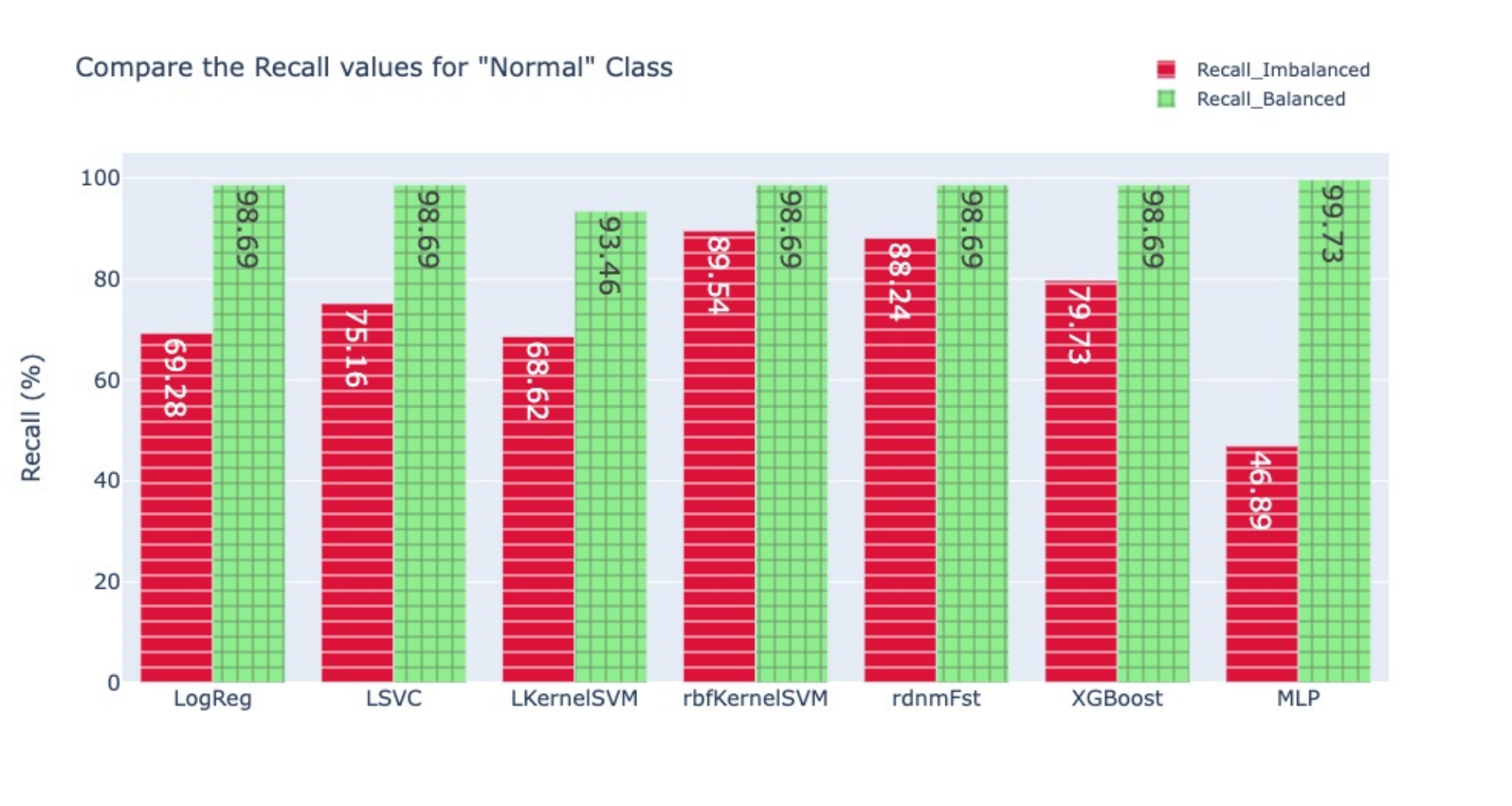}
                \subcaption{Recall on Imbalanced and Balanced data.}
                \label{fig:Recall}
        \end{subfigure}%
        \hfill
        \begin{subfigure}[!htb]{0.8\textwidth}
            \centering
                \includegraphics[width=\textwidth]{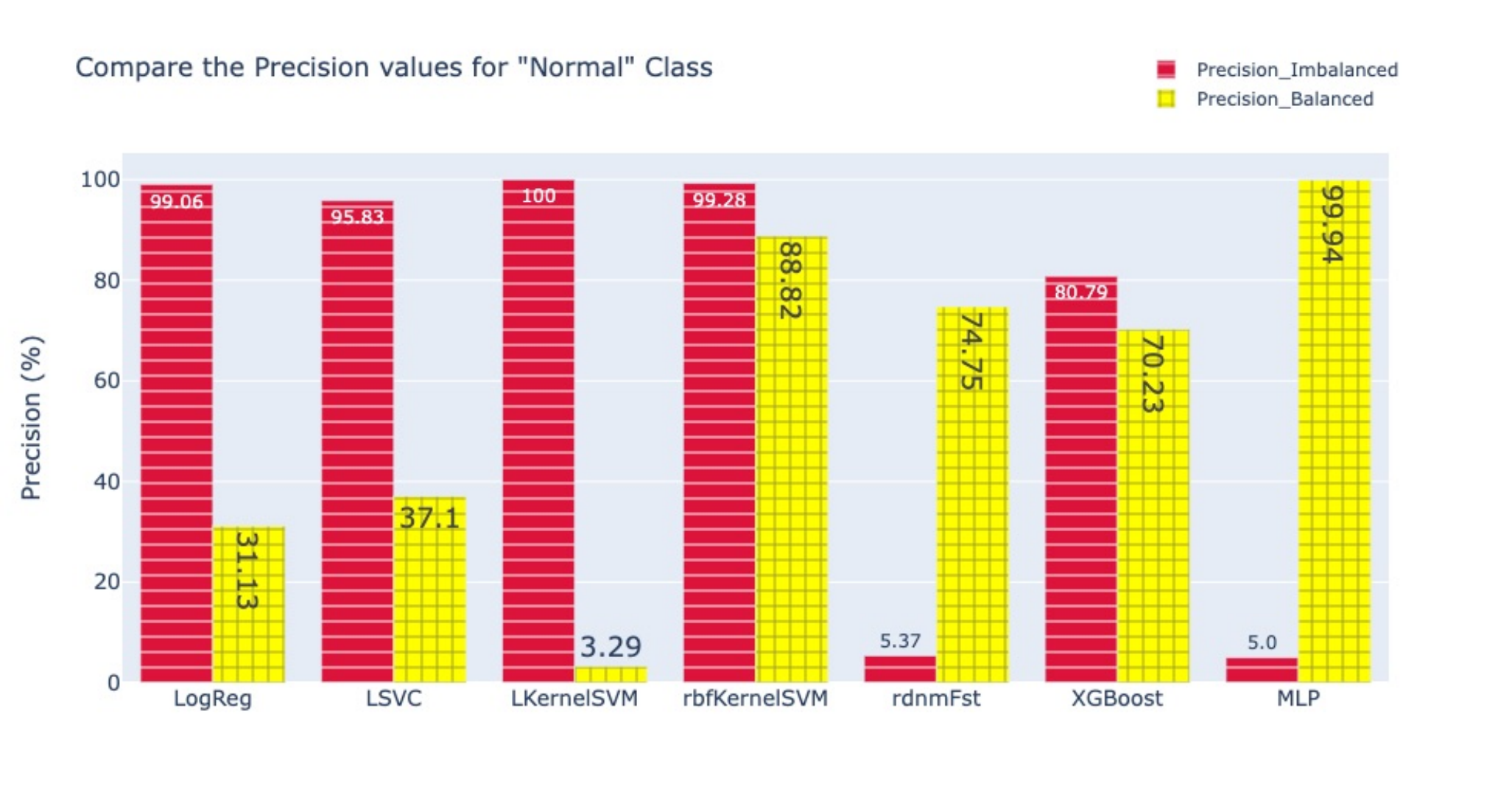}
                \subcaption{Precision on Imbalanced and Balanced data.}
                \label{fig:Precision}
        \end{subfigure}%
    \caption{SMOTE impact on Precision and Recall of Minority Class.}
\end{figure*}

\begin{figure*}[!htb]
    \centering
    \includegraphics[width=0.7\textwidth]{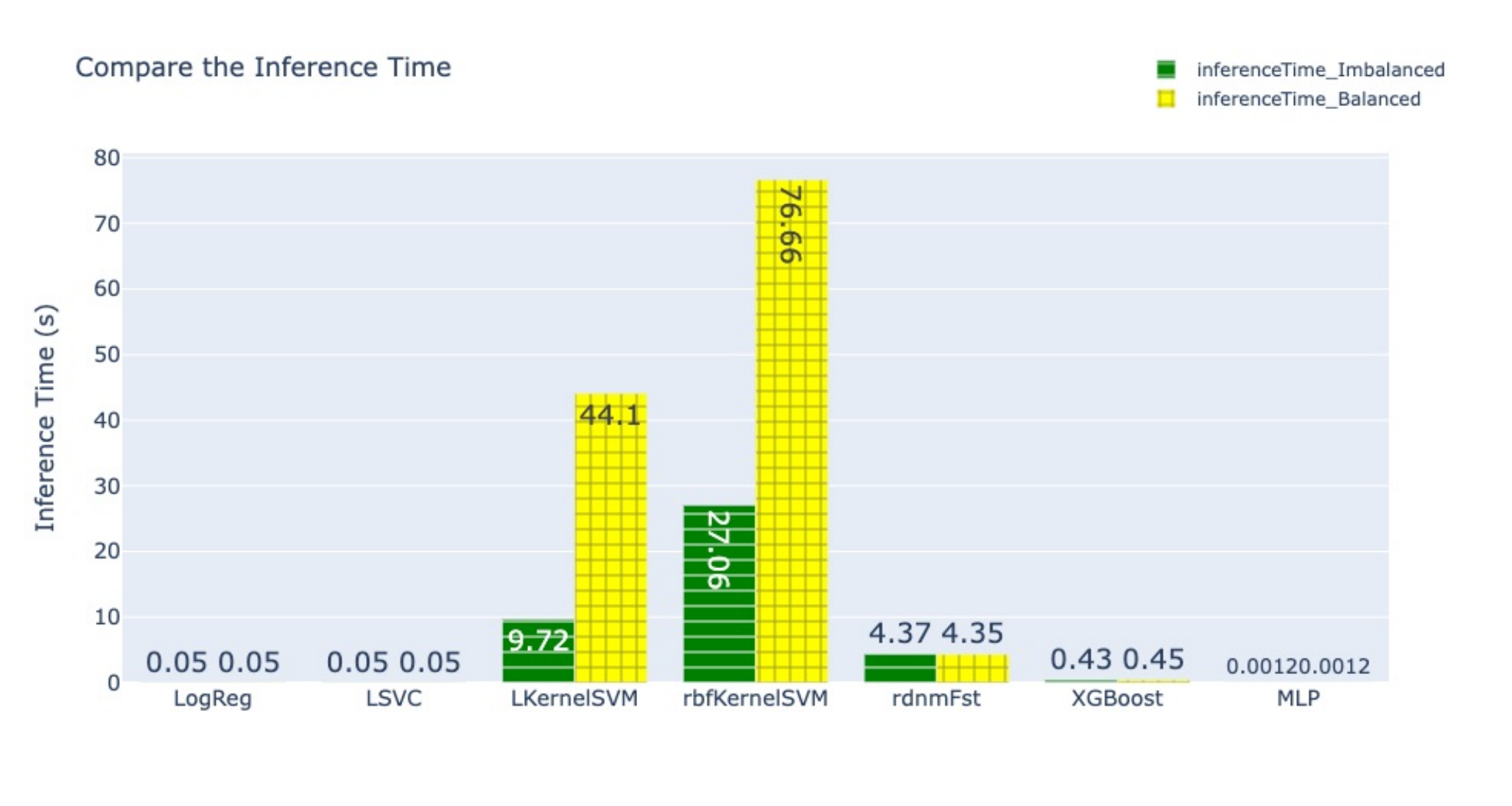}
    \caption{Inference Time on Imbalanced and Balanced data.}
    \label{fig:InferenceTime}
\end{figure*}

Figures~\ref{fig:FPR},\ref{fig:FNR},\ref{fig:InferenceTime},\ref{fig:Recall},\ref{fig:Precision} that follow illustrate the observations observe. In these figures; LogReg, LSVC, LKernelSVM, rbfKernelSVM, rdnmFst, XGBoost, and MLP refer to Logistic Regression, Linear SVC, Linear Kernel SVM, RBF Kernel SVM, Random Forest, Extreme Gradient Boosting, and Multi-layer Perceptron respectively.

\newpage
\section{Conclusion}
\label{sec:Conclusion}
The existing Bot-IoT dataset contains imbalanced normal and attack packets since the number of normal packets is much smaller than that of attack ones.
Such class imbalance leads to inaccurate results, especially for the minority class.
In this thesis, we have addressed the class imbalance problem to apply the SMOTE algorithm to the original Bot-IoT dataset, where the SMOTE algorithm generates synthetic samples such that number of normal packets is equivalent to that of attack ones.
We further have proposed the binary classification method to identify the normal and attack packet.

Through the numerical results, we first have shown the performance comparison among the different classifiers.
We observe that all the proposed classifiers achieve high accuracy, recall, and precision.
Next, we have demonstrated the impact of data sampling on the classifier performance.
The proposed classifiers trained in the balanced dataset also achieves almost the same accuracy, recall, and precision as that in imbalanced dataset, and drastically improves the FPR.

In future work, we plan to develop the resource efficient NIDS based on both the reinforcement learning and federated learning to learn patterns of attacks in the IoT network over time.
In addition, we aim to evaluate the performance of the NIDS under the resource constraints. The more immediate efforts include extending this work to do multi-class classification especially when an attack has been detected, in order to identify the actual type of attack. 
\newpage
\section*{Acknowledgment}
I extend my sincere appreciation to my principal supervisor Professor Shoji Kasahara, for his continued guidance over the master's course. Our time together at the Large-scale Systems Management Laboratory has certainly shaped me into the researcher I am today. It is due to his mentorship that I have been able to undertake such a challenging but very satisfying research project. I will always be indebted to him and I am very grateful for the opportunity he gave me to join his Laboratory at Nara Institute of Science and Technology (NAIST). He's made my life in Japan convenient for research such that I have not been stressed throughout graduate school. Moreover, I am very thankful for the permission and the opportunity he gave me to undertake an internship with HONDA Research Institute Japan (HRI-JP). This has broadened my knowledge and understanding of specific Artificial Intelligence disciplines especially Robotics and Natural Language Processing. Sensei, I will always be grateful.  

I’d also like to thank all my co-supervisors: Professor Youki Kadobayashi, Associate Professor Masahiro Sasabe, Associate Professor YuanYu Zhang, and Assistant Professor Takanori Hara, for the relentless support and guidance throughout this research project on Network Intrusion Detection.  I am really delighted with the insights on the research direction that I got because I couldn't make it to the end without such valuable help.
Assistant Professor Takanori Hara has been very resourceful to me in terms of setting up the coding environments required to conduct the research experiments and the revision of this thesis. I appreciate your tremendous help Dr. Hara.

I cannot forget the valuable time I spent with  Professor Masatoshi Yoshikawa at the Graduate school of Informatics of Kyoto University. His daily encouragement and words of advice helped me a lot to master several methods and techniques in Information Retrieval, Databases, Human-Computer Interface design and Artificial Intelligence.

I want to extend my appreciation to Kentaro Fujita who I met at the Large-scale Systems Management Laboratory. He played such a fundamental role in showing me how to implement several machine learning algorithms in Python. Thank you Fujita-san.

My friends at the laboratory have always been there for me, even as we have been physically distanced by the COVID-19 pandemic. I started my research in the middle of the pandemic and only the constant updates from my friends—at the lab and the school at large—helped me to keep up with the latest information regarding laboratory research meetings, and seminars. Thank you

A big reason for my success has been the incredibly selfless love I have always gotten from my parents. To you, mom and dad, I say thank you very much. I extend special thanks to my siblings with whom we have continued to challenge each other to do better every day. To everyone back in Wakiso, Uganda and beyond, I am forever grateful for the incredible moments we have always shared together.

\newpage
\bibliographystyle{plain} 
\bibliography{mthesis}

\newpage
\section*{Publication List}
\addcontentsline{toc}{section}{Publication List}
\begin{enumerate}
    \item Jesse Atuhurra, Takanori Hara, Yuanyu Zhang and Shoji Kasahara, ``On Attack Pattern Classification in IoT Networks for Network Intrusion Detection Systems,'' IEICE Tech. Rep., Nov. 2021.
\end{enumerate}

\end{document}